\documentclass[12pt]{article}
\usepackage{graphicx,amsmath,amssymb}

\newlength{\dinwidth}
\newlength{\dinmargin}
\newcommand{\dif}{\mathrm{d}}

\newcommand{\xB}{x_{\scriptscriptstyle{B}}}
\newcommand{\Pom}{{\hspace{ -0.1em}I\hspace{-0.2em}P}}
\newcommand{\Reg}{{\hspace{ -0.1em}I\hspace{-0.2em}R}}
\newcommand{\xPom}{x_\Pom}
\newcommand{\chisq}{\chi^2/\mathrm{d.o.f.}}

\begin{document}
\titlepage
\begin{flushright}
  IPPP/04/09     \\
  DCPT/04/18     \\
  9th November 2004 \\
\end{flushright}

\vspace*{0.5cm}

\begin{center}
  
  {\Large \bf A QCD analysis of diffractive\\[1ex] deep-inelastic scattering data}

  \vspace*{1cm}

  \textsc{A.D. Martin$^a$, M.G. Ryskin$^{a,b}$ and G. Watt$^a$} \\

  \vspace*{0.5cm}

  $^a$ Institute for Particle Physics Phenomenology, University of Durham, DH1 3LE, UK \\
  $^b$ Petersburg Nuclear Physics Institute, Gatchina, St.~Petersburg, 188300, Russia

\end{center}

\vspace*{0.5cm}

\begin{abstract}
  We perform a novel type of analysis of diffractive deep-inelastic scattering data, in which the input parton distributions of the Pomeron are parameterised using the perturbative QCD expressions.  In particular, we treat individually the components of the Pomeron of different size.  We are able to describe simultaneously both the recent ZEUS and H1 diffractive data.  In addition to the usual two-gluon model for the perturbative Pomeron, we allow for the possibility that it may be made from two sea quarks.
\end{abstract}

A notable feature of deep-inelastic scattering is the existence of diffractive events, $\gamma^* p\to X p$, in which the slightly deflected proton and the cluster $X$ of outgoing hadrons are well-separated in rapidity.  The large rapidity gap is believed to be associated with Pomeron exchange.  The diffractive events make up an appreciable fraction of all (inclusive) deep-inelastic events, $\gamma^* p \to X$.  We will refer to the diffractive and inclusive processes as DDIS and DIS respectively.

Here we perform a perturbative QCD analysis of the new high precision DDIS data, recently obtained by the ZEUS \cite{ZEUSLPS,ZEUSMX} and H1 \cite{H1data} Collaborations at HERA.   The analysis is novel in that it treats individually the components of the Pomeron of different transverse size.  The description of the DDIS data is based on a purely perturbative QCD framework.  We take input forms of the parton distributions of the Pomeron given by the calculation of the lowest-order QCD diagrams for $\gamma^* p\to X p$ \cite{Wusthoff:1997fz}.  In previous analyses, the Pomeron was treated as a hadron-like object of more or less fixed size.  However, the microscopic structure of the Pomeron is different to that of a hadron.  In perturbative QCD, it is known that the Pomeron singularity is not an isolated pole, but a branch cut, in the complex angular momentum plane \cite{Lipatov:1985uk}.  The pole singularity corresponds to a single particle, whereas a branch cut may be regarded as a continuum series of poles.  That is, the Pomeron wave function consists of a continuous number of components.  Each component $i$ has its own size, $1/\mu_i$.  The QCD DGLAP evolution of a component should start from its own scale $\mu_i$, provided that $\mu_i$ is large enough for the perturbative evolution to be valid.  Therefore, the expression for the diffractive structure function $F_2^D$ contains an integral over the Pomeron size, or rather over the scale $\mu$.  So to obtain $F_2^D$ we evolve the input parton distributions of each component of the Pomeron from their own starting scale $\mu$ up to the final scale $Q$.  The extra integral over $\mu$ reflects the fact that the partonic structure of the Pomeron is more complicated than that of a normal hadron.

\begin{figure}
  \centering
  \begin{minipage}{0.49\textwidth}
    \hspace{0.35\textwidth}(a)\\
    \includegraphics[width=0.9\textwidth]{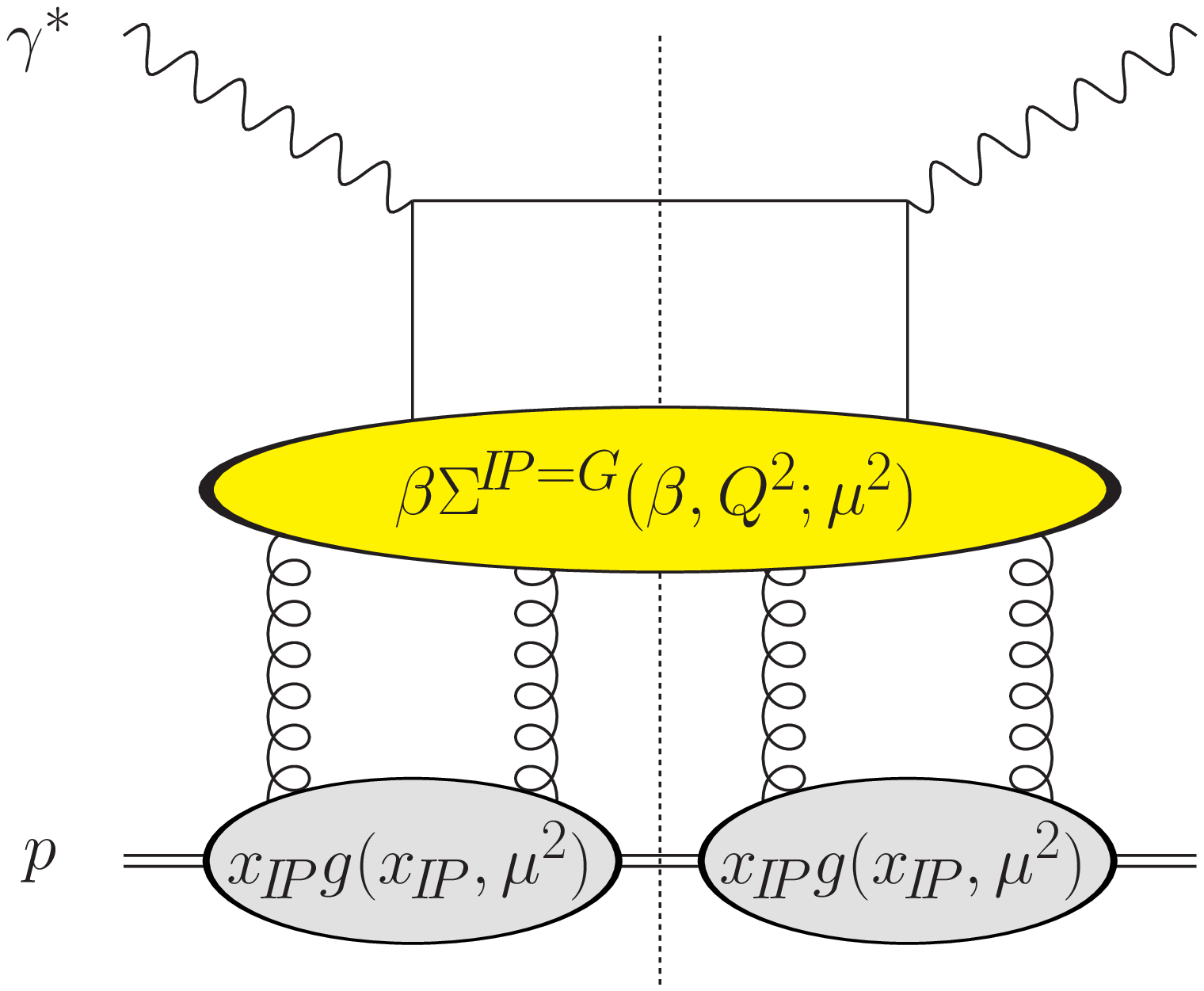}\hfill
  \end{minipage}
  \begin{minipage}{0.49\textwidth}
    \hspace{0.35\textwidth}(b)\\
    \hfill\includegraphics[width=0.9\textwidth]{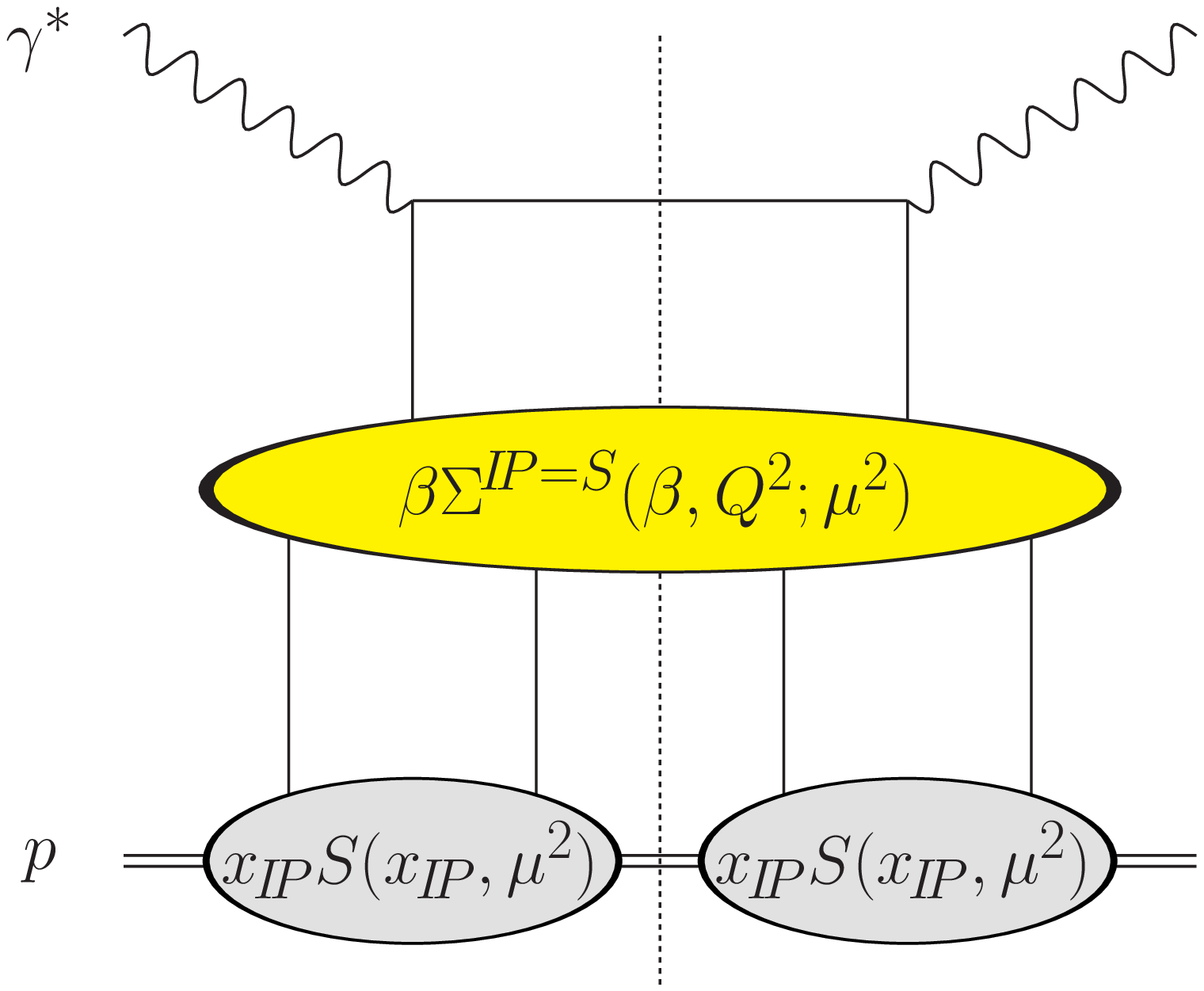}
  \end{minipage}
  \caption{(a) Cut diagram illustrating the main idea contained in \eqref{eq:F2D3P}.  For each component of the perturbative Pomeron of size $1/\mu$, represented by two $t$-channel gluons in a colour singlet, the Pomeron structure function $F_{2}^{\Pom}(\beta,Q^2;\mu^2)$ is evaluated from the quark singlet, $\beta \Sigma^\Pom(\beta,Q^2;\mu^2)$, and gluon, $\beta g^\Pom(\beta,Q^2;\mu^2)$, distributions of the Pomeron.  The perturbative Pomeron flux factor $f_{\Pom}(\xPom;\mu^2)$ is given in terms of the gluon distribution of the proton, $\xPom\,g(\xPom,\mu^2)$.  (b) Later, we also include contributions from diagrams in which the Pomeron is represented by sea quark--antiquark exchange (plus interference with the two-gluon Pomeron).}
  \label{fig:f2d3pom}
\end{figure}

Recall that in the usual analyses (for example, by H1 \cite{H1data}) Regge factorisation \cite{Ingelman:1984ns} is assumed, such that the diffractive structure function $F_2^{D(4)}(\xPom,\beta,Q^2,t)$\footnote{Here, $\xPom$ is the fraction of the proton's momentum transferred through the rapidity gap by the Pomeron, $\beta\equiv \xB/\xPom$ is the fraction of the Pomeron's momentum carried by the struck quark, $\xB$ is the Bjorken $x$ variable, $Q^2$ is the photon virtuality, and $t$ is the squared 4-momentum transfer.} is written as a product of the Pomeron flux factor $f_\Pom(\xPom,t)$ and the structure function $F_2^\Pom(\beta,Q^2)$ which describes the interaction of the Pomeron with the virtual photon probe.  The input Pomeron parton distributions are taken to be arbitrary polynomials.  The Pomeron flux factor is taken from Regge phenomenology with some effective Pomeron intercept $\alpha_\Pom(0)$.  However, the value of $\alpha_\Pom(0)=1.17$ \cite{H1data} needed to fit DDIS data is significantly higher than the value of 1.08 obtained from soft hadron data \cite{Donnachie:1992ny}.  Instead, in the present analysis, we have $\mu$-factorisation, such that the $t$-integrated observable is 
\begin{equation}
  \label{eq:F2D3P}
  F_{2,{\rm P}}^{D(3)}(\xPom,\beta,Q^2) = \int_{Q_0^2}^{Q^2}\dif\mu^2\;f_{\Pom}(\xPom;\mu^2)\;F_{2}^\Pom(\beta,Q^2;\mu^2);
\end{equation}
see Fig.~\ref{fig:f2d3pom}.  Here, the subscript ${\rm P}$ on $F_{2,{\rm P}}^{D(3)}$ is to indicate that this is the perturbative contribution with $\mu>Q_0\sim 1$ GeV.
\begin{figure}
  \centering
  \begin{minipage}{0.49\textwidth}
    \hspace{0.5\textwidth}(a)\\
    \includegraphics[width=\textwidth]{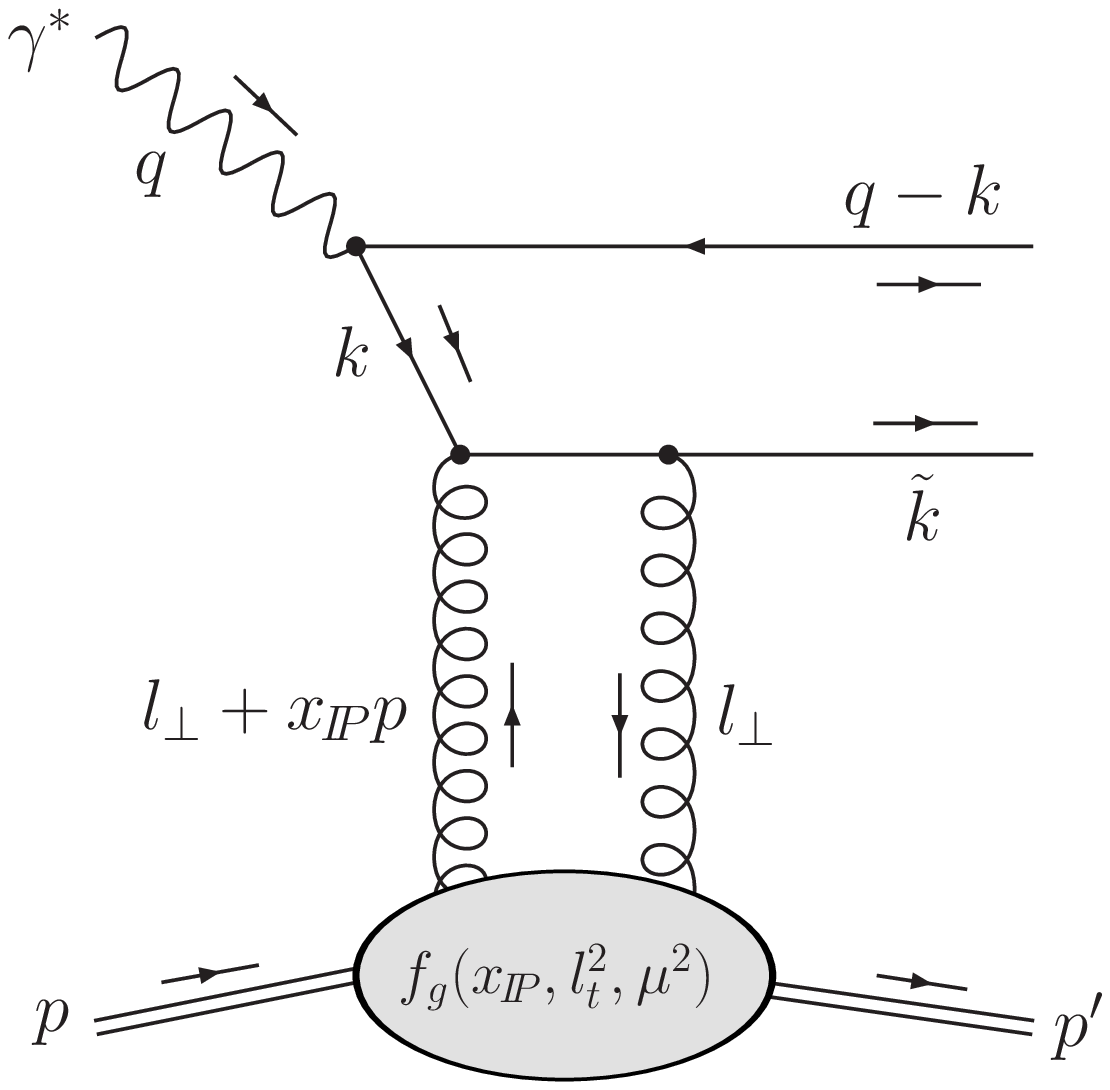}
  \end{minipage}
  \begin{minipage}{0.49\textwidth}
    \hspace{0.5\textwidth}(b)\\
    \includegraphics[width=\textwidth]{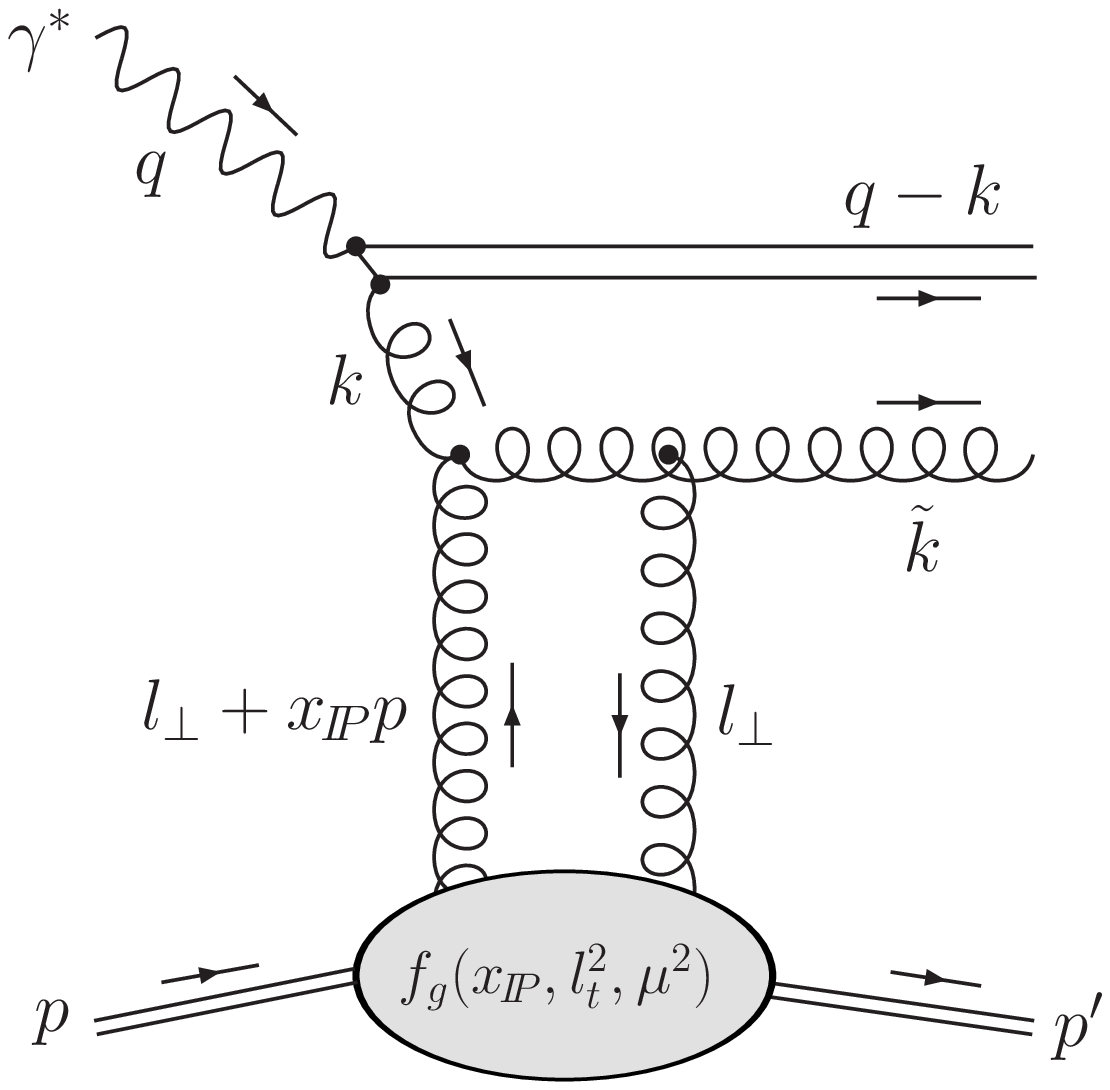}
  \end{minipage}
  \caption{(a) Quark dipole and (b) effective gluon dipole interacting with the proton via a perturbative Pomeron composed of two $t$-channel gluons.  Here, $l_\perp$ is a space-like 4-vector such that $l_\perp^2=-l_t^2$, and $f_g(\xPom,l_t^2,\mu^2)$ is the unintegrated gluon distribution of the proton.}
  \label{fig:dipoleG}
\end{figure}
We will introduce the full form which we take for $F_2^{D(3)}$ in a moment.

The Pomeron structure function, $F_{2}^\Pom(\beta,Q^2;\mu^2)$, is obtained by NLO DGLAP evolution up to $Q^2$ from input Pomeron parton distributions parameterised at a starting scale $\mu^2$.  The $\beta$ dependence of these input distributions is obtained from lowest-order perturbative QCD diagrams.  The perturbative Pomeron is represented by two $t$-channel gluons in a colour singlet; we will indicate this using the notation $\Pom=G$.  The relevant lowest-order diagrams are shown in Fig.~\ref{fig:dipoleG}.  We regard the virtual photon as dissociating into either a quark--antiquark dipole (Fig.~\ref{fig:dipoleG}(a)) or an effective gluon dipole, made up of a gluon and a compact $q\bar{q}$ pair (Fig.~\ref{fig:dipoleG}(b)).  In each case the four permutations of the couplings of the two $t$-channel gluons to the two components of the dipole are implied.  Evaluating these diagrams using light-cone wave functions of the photon, in the limit of strongly-ordered transverse momenta ($l_t\ll k_t\ll Q$), leads to the input quark singlet and gluon distributions of the form \cite{Wusthoff:1997fz}
\begin{align}
  \beta \Sigma^{\Pom=G}(\beta,\mu^2; \mu^2) &= c_{q/G}\,\beta^3\,(1-\beta), \label{eq:dpdfqpG} \\
  \beta^\prime g^{\Pom=G}(\beta^\prime,\mu^2; \mu^2) &= c_{g/G}\,(1+2\beta^\prime)^2\,(1-\beta^\prime)^2, \label{eq:dpdfgpG}
  \end{align}
where $\beta^\prime$ is the Pomeron's momentum fraction carried by the gluon with momentum $k$ in Fig.~\ref{fig:dipoleG}(b).  The parameters $c_{q/G}$ and $c_{g/G}$ implicitly include all the numerical factors arising from the lowest-order calculations.  We will let these normalisations go free in fits to the DDIS data to account for higher-order QCD corrections (effective $K$-factors).  The quark singlet distribution is $\Sigma^{\Pom}\equiv u+d+s+\bar{u}+\bar{d}+\bar{s}$, with $u=d=s=\bar{u}=\bar{d}=\bar{s}$, so that the non-singlet distributions are all zero.  The contributions of the charm and bottom quarks to $F_{2}^{\Pom}(\beta,Q^2;\mu^2)$ are calculated in the heavy quark fixed-flavour number scheme.

Moreover, for the perturbative contribution, the Pomeron flux factor $f_{\Pom}(\xPom;\mu^2)$ is given in terms of the integrated gluon distribution of the proton. Consider the lowest-order diagram, Fig.~\ref{fig:dipoleG}(a), in which the virtual photon dissociates into a forward-going quark--antiquark pair, that is, $\gamma^*\to q\bar{q}$ as $t\to 0$ \cite{Wusthoff:1997fz,Nikolaev:1990ja}.  After summation over the four amplitudes, the integral of the unintegrated gluon distribution $f_g(\xPom,l_t^2,\mu^2)$ over the gluon loop transverse momentum $l_t^2$ up to the quark virtuality $\mu^2 = k_t^2/(1-\beta)$ gives the integrated gluon distribution $\xPom\,g(\xPom,\mu^2)$.  Thus the Pomeron flux factor is
\begin{equation}
  f_{\Pom=G}(\xPom;\mu^2) = \frac{1}{\xPom} \left[\,\frac{\alpha_S(\mu^2)}{\mu^2}\;\xPom\,g(\xPom,\mu^2)\,\right]^2.
\label{eq:PpomfluxG}
\end{equation}
Here, a factor $1/B_D$ from the $t$-integration, where $B_D\simeq 6$ GeV$^{-2}$ is the diffractive slope parameter, has been absorbed into the parameters $c_{q/G}$ and $c_{g/G}$ of \eqref{eq:dpdfqpG} and \eqref{eq:dpdfgpG}.  Strictly speaking, the Pomeron flux factor \eqref{eq:PpomfluxG} should be written in terms of the \emph{skewed} gluon distribution.  At small $\xPom$ this gives rise to an overall constant factor \cite{Shuvaev:1999ce}, $R_g^2$, which again we absorb into the parameters $c_{q/G}$ and $c_{g/G}$.

In addition to the leading-twist contribution arising from Fig.~\ref{fig:dipoleG}(a) with a transversely polarised photon, there is an analogous twist-four contribution to $F_2^{D(3)}$ arising from Fig.~\ref{fig:dipoleG}(a) with a longitudinally polarised photon,
\begin{equation}
  \label{eq:FLD3P}
  F_{L,{\rm P}}^{D(3)}(\xPom,\beta,Q^2) = \left(\int_{Q_0^2}^{Q^2}\dif\mu^2\;\frac{\mu^2}{Q^2}\;f_{\Pom}(\xPom;\mu^2)\right)\;F_L^{\Pom}(\beta).
\end{equation}
The $\beta$ dependence is again obtained from lowest-order perturbative QCD calculations \cite{Wusthoff:1997fz}:
\begin{equation} \label{eq:FLbeta}
  F_{L}^{\Pom=G}(\beta) = c_{L/G}\;\beta^3\,(2\beta-1)^2,
\end{equation}
where, as before, $c_{L/G}$ is taken to be a free parameter.  The twist-four nature of this longitudinal contribution is evident from the $\mu^2/Q^2$ factor in \eqref{eq:FLD3P}.

We also include a non-perturbative (${\rm NP}$) Pomeron contribution (from scales $\mu<Q_0$) and a secondary Reggeon ($\Reg$) contribution to $F_2^{D(3)}(\xPom,\beta,Q^2)$, so that
\begin{equation}
  \label{eq:F2D3sum}
  F_2^{D(3)} = F_{2,{\rm P}}^{D(3)} + F_{2,{\rm NP}}^{D(3)} + F_{L,{\rm P}}^{D(3)} + F_{2,\Reg}^{D(3)},
\end{equation}
with
\begin{equation}
  \label{eq:F2D3NP}
  F_{2,{\rm NP}}^{D(3)}(\xPom,\beta,Q^2) = f_{\Pom={\rm NP}}(\xPom)\,F_{2}^{\Pom={\rm NP}}(\beta,Q^2;Q_0^2),
\end{equation}
\begin{equation}
  \label{eq:F2D3R}
  F_{2,\Reg}^{D(3)}(\xPom,\beta,Q^2) = c_\Reg\,f_\Reg(\xPom)\,F_2^\Reg(\beta,Q^2),
\end{equation}
where $c_\Reg$ is taken to be a free parameter.  Here, the non-perturbative Pomeron and Reggeon flux factors are\footnote{The couplings of the Pomeron or secondary Reggeon to the proton are absorbed into the parameters $c_{a/{\rm NP}}$ ($a=q,g$) and $c_\Reg$.}
\begin{equation}
  \label{eq:NPregflux}
  f_{i}(\xPom) = \int_{t_{\mathrm{cut}}}^{t_{\mathrm{min}}}\!\dif{t}\;\frac{\mathrm{e}^{B_i\,t}}{\xPom^{2\alpha_i(t)-1}} = \frac{\xPom^{1-2\alpha_i(0)}\left(1-\mathrm{e}^{B_i\,t_{\mathrm{cut}}}\xPom^{-2\alpha_i^\prime\,t_{\mathrm{cut}}}\right)}{B_i+2\alpha_i^\prime \ln(1/\xPom)},
\end{equation}
with $i=\Pom$ and $\Reg$ respectively, and $\alpha_i(t) = \alpha_i(0)+\alpha_i^\prime\,t$.  The integration limits are taken to be $t_{\mathrm{cut}}=-1$ GeV$^2$ and $t_{\mathrm{min}}\approx 0$ GeV$^2$.  For the non-perturbative Pomeron, we fix $\alpha_\Pom(0)=1.08$ \cite{Donnachie:1992ny}, $\alpha_\Pom^\prime=0.26\,\mathrm{GeV}^{-2}$, and $B_\Pom=4.6\,\mathrm{GeV}^{-2}$ \cite{Abe:1993xx}, whereas for the secondary Reggeon we take $\alpha_\Reg(0)=0.50$ \cite{Adloff:1997sc}, $\alpha_\Reg^\prime=0.90\,\mathrm{GeV}^{-2}$ \cite{Apel:1979sp}, and $B_\Reg=2.0\,\mathrm{GeV}^{-2}$ \cite{Kaidalov:jz}.  Apart from $\alpha_\Pom(0)$, these are the same values used in the preliminary H1 analysis \cite{H1data}.  The secondary Reggeon structure function, $F_2^\Reg(\beta,Q^2)$, is calculated at NLO from the GRV pionic parton distributions \cite{Gluck:1991ey}.  For the non-perturbative Pomeron, the input quark singlet and gluon distributions, $\beta \Sigma^{\Pom={\rm NP}}(\beta,Q_0^2; Q_0^2)$ and $\beta^\prime g^{\Pom={\rm NP}}(\beta^\prime,Q_0^2; Q_0^2)$, are taken to have the same $\beta$ dependence as for the two-quark Pomeron introduced later on (see \eqref{eq:dpdfqpS} and \eqref{eq:dpdfgpS}), with different normalisations $c_{q/{\rm NP}}$ and $c_{g/{\rm NP}}$.  (Taking the same $\beta$ dependence as for the two-gluon Pomeron, $\eqref{eq:dpdfqpG}$ and $\eqref{eq:dpdfgpG}$, gives a much worse description of the data.)

We fit to the preliminary ZEUS \cite{ZEUSLPS,ZEUSMX} and H1 \cite{H1data} DDIS data using \eqref{eq:F2D3sum}, and varying the free parameters until an optimum description of the data is obtained.  We impose a cut $M_X>2$ GeV on the fitted data to exclude large contributions from vector meson production and other higher-twist effects, and a cut $y<0.45$ so that we can assume that the measured reduced diffractive cross section $\sigma_r^{D(3)}$ is approximately equal to $F_2^{D(3)}$.  The statistical and systematic experimental errors are added in quadrature.  We use the \textsc{qcdnum} program \cite{QCDNUM} to perform the NLO DGLAP evolution and the \textsc{minuit} program \cite{James:1975dr} to find the optimal parameters.  The values of $\alpha_S(M_Z^2)$ and the charm and bottom quark masses are taken to be the same as in the MRST2001 NLO parton set \cite{MRST2001}.  Two sets of preliminary ZEUS data are fitted: those obtained using the leading proton spectrometer (LPS) \cite{ZEUSLPS}, and those obtained using the so-called $M_X$ method \cite{ZEUSMX} which is based on the fact that diffractive and non-diffractive events have very different $\ln M_X^2$ distributions.  For the latter data set, in addition to elastic proton scattering, proton dissociation up to mass $M_Y=2.3$ GeV is included.  Clearly the cross section will be larger in this case, so we allow for the overall normalisation of these data by multiplying \eqref{eq:F2D3sum} by a factor $N_Z$.   An analogous normalisation, $N_H$, is applied for the preliminary H1 data \cite{H1data}, where diffractive events are selected on the basis of a large rapidity gap, and where proton dissociation up to mass $M_Y=1.6$ GeV is included.  The ZEUS $M_X$ data \cite{ZEUSMX} do not include secondary Reggeon contributions, therefore we omit the fourth term of \eqref{eq:F2D3sum} when fitting to these data.  We fit to each data set separately, and then we perform fits to the three data sets combined.

For our first study, we parameterise the perturbative Pomeron flux factor \eqref{eq:PpomfluxG} using a simplified form for the gluon distribution of the proton,
\begin{equation}
  \label{eq:lambda}
  \xPom g(\xPom,\mu^2)=\xPom^{-\lambda},
\end{equation}
where $\lambda$ is independent of $\mu^2$ and is determined by the fit to data.\footnote{Strictly speaking, $\lambda$ should depend on $\ln\mu^2$.  We investigated this effect by taking $\lambda(\mu^2) = 0.08 + c_\lambda\ln(\mu^2/(0.45\,\mathrm{GeV}^2))$ with $Q_0=1$ GeV and $c_{g/{\rm NP}}=0$.  The combined fit to ZEUS and H1 DDIS data gave a $\chisq=1.12$ with $c_\lambda = 0.054\pm0.006$.  This is consistent with the value found by H1 in a fit to inclusive $F_2$ data \cite{Adloff:2001rw} of $c_\lambda=0.0481\pm0.0013(\mathrm{stat.})\pm0.0037(\mathrm{syst.})$.  Since the $\chisq$ was not improved compared to the corresponding fit which took $\lambda$ to be independent of $\mu^2$ ($\chisq=1.07$), we used the form \eqref{eq:lambda} for simplicity.}  The normalisation of \eqref{eq:lambda} has been absorbed into the free parameters $c_{q/G}$, $c_{g/G}$, and $c_{L/G}$.

Varying the $Q_0$ parameter, we find that the best fit to the combined ZEUS and H1 data sets is obtained with $Q_0^2 = 0.8$ GeV$^2$, which gives a $\chisq=1.05$ with $c_{g/{\rm NP}}$ going to zero.  Later on, we will use the MRST2001 NLO \cite{MRST2001} parton distributions of the proton instead of the simplified form \eqref{eq:lambda}, where the minimum possible scale is 1 GeV.  Using the form \eqref{eq:lambda} with $Q_0^2=1$ GeV$^2$ gives only a slightly worse $\chisq=1.07$.  Furthermore, fixing $c_{g/{\rm NP}} = 0$ makes little difference to the quality of the fit.  Therefore, in all fits presented here, we take $Q_0^2=1$ GeV$^2$ and fix $c_{g/{\rm NP}} = 0$.

\begin{figure}
  \centering
  \includegraphics[width=0.5\textwidth]{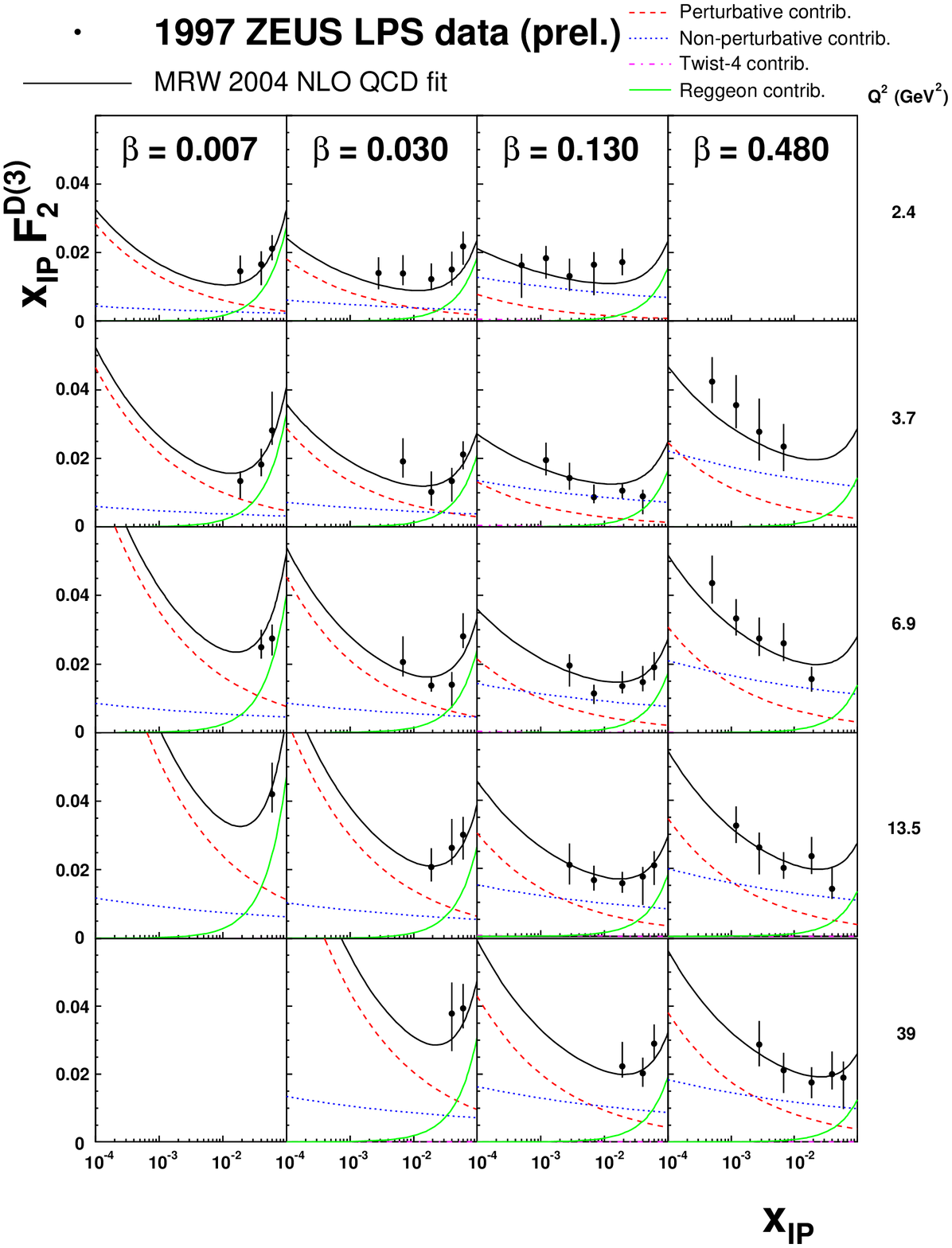}\hfill%
  \includegraphics[width=0.5\textwidth]{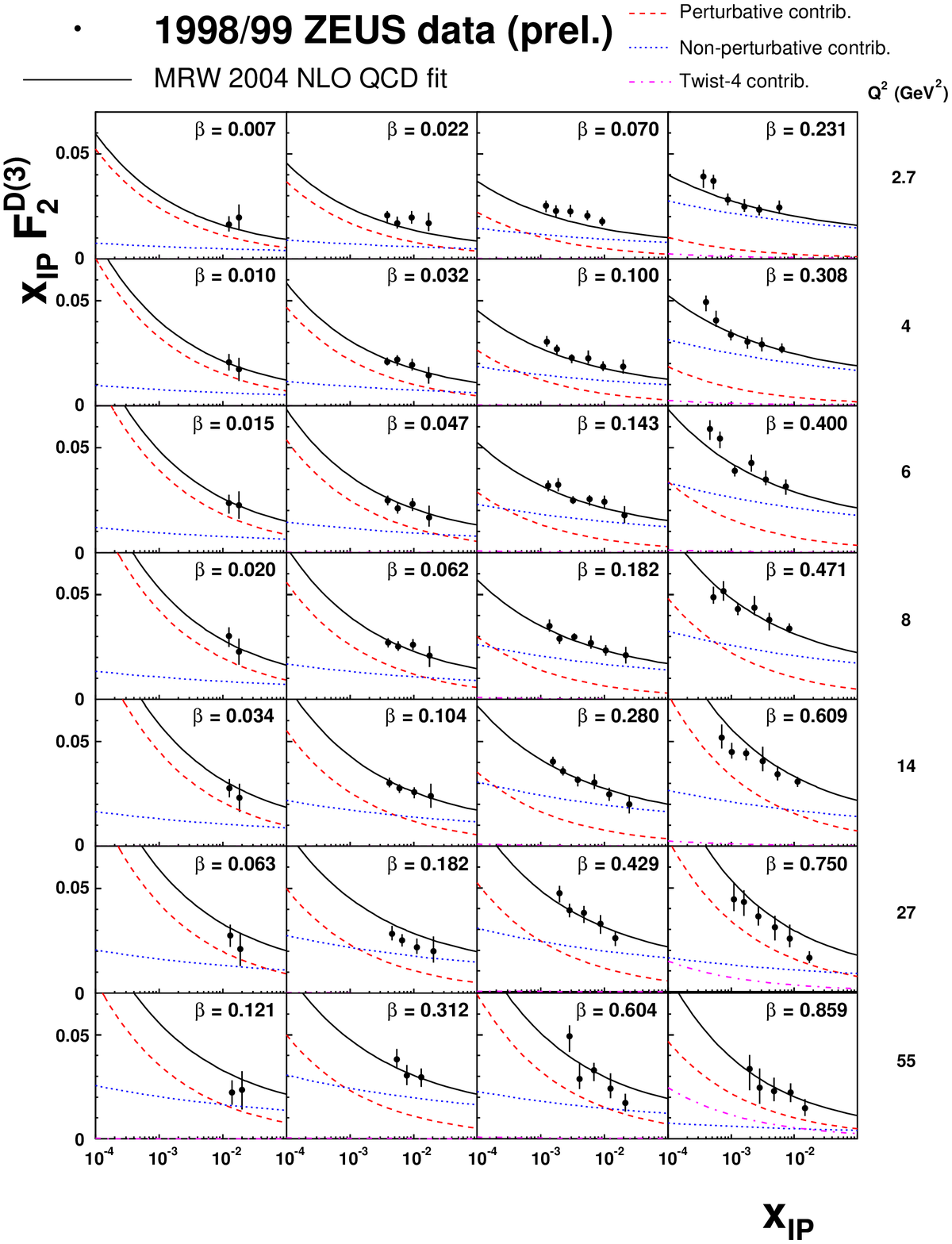}\\
  \includegraphics[width=0.9\textwidth]{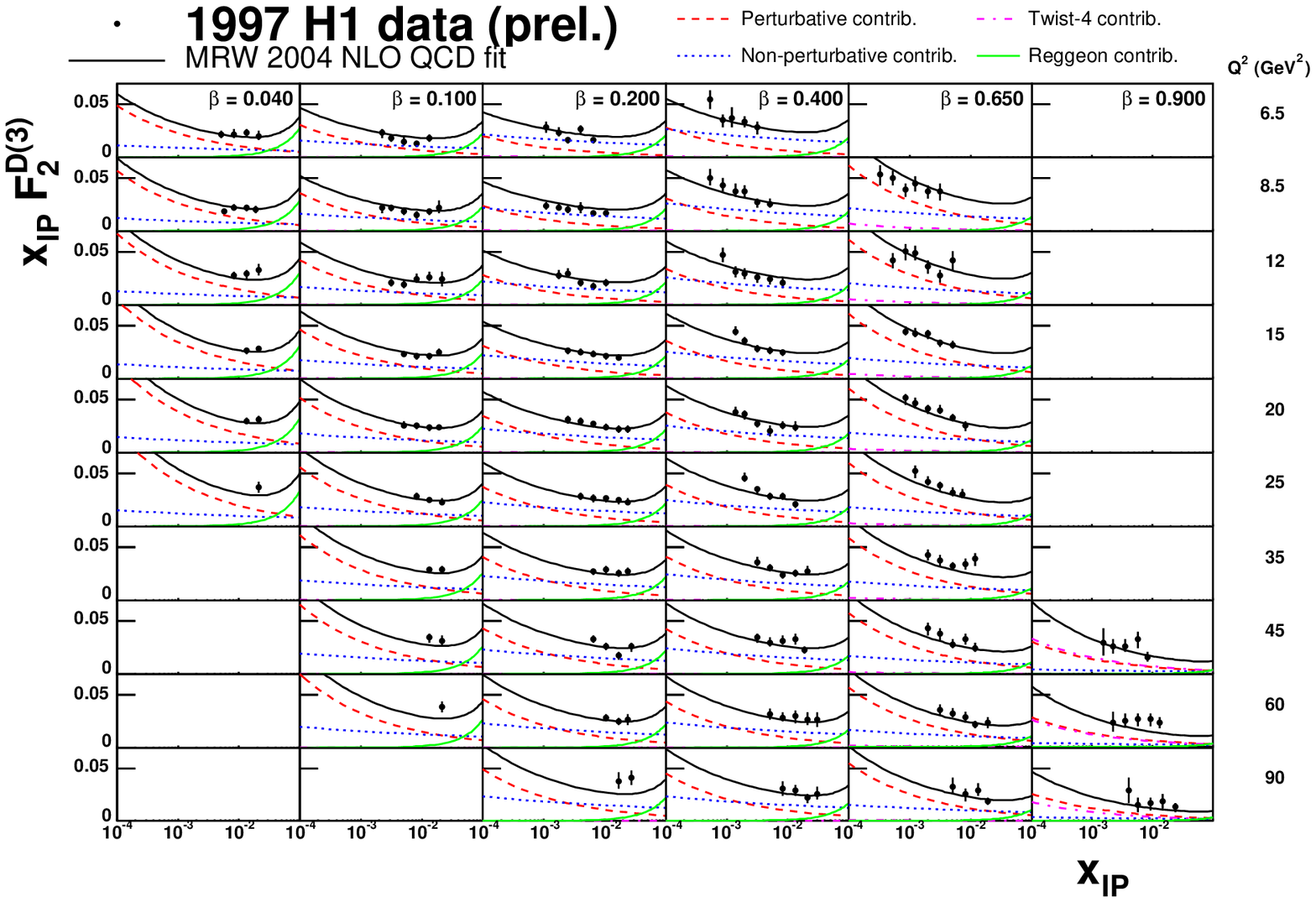}
  \caption{Fit to combined preliminary ZEUS \cite{ZEUSLPS,ZEUSMX} and H1 \cite{H1data} $F_2^{D(3)}$ data with a gluon distribution of the proton proportional to $\xPom^{-\lambda}$ \eqref{eq:lambda}.  The curves show the four contributions to the total, as defined in \eqref{eq:F2D3sum}.  Only data points included in the fit are plotted.}
  \label{fig:dummyG}
\end{figure}

We find that each data set can be well described by this simple, perturbatively-motivated, approach.   However, different values of $\lambda$ and the other parameters are obtained from the ZEUS and H1 data, as can be seen from Table \ref{tab:dummyG}.  In particular, the H1 data seem to have a flatter $\xPom$ dependence than the ZEUS data.  This should be regarded as some inconsistency between the data sets, but not as a contradiction, since it is possible to obtain an adequate description of the combined data sets, as shown in Fig.~\ref{fig:dummyG} and by the results in the last column of Table \ref{tab:dummyG}.

\begin{table}
  \centering
  \begin{tabular}[c]{lcccc}
    \hline
    Data sets fitted & ZEUS LPS & ZEUS $M_X$ & H1 & ZEUS + H1 \\
    Number of points & 69 & 121 & 214 & 404 \\
    \hline
    $\chisq$ & $0.67$ & $0.78$ & $1.08$ & $1.08$ \\[2mm]\hline
    $c_{q/G}$ (GeV$^2$) & $0.71\pm0.39$ & $0.48\pm0.12$ & $2.2\pm0.4$ & $1.13\pm0.15$ \\[2mm]
    $c_{g/G}$ (GeV$^2$) & $0.11\pm0.05$ & $0.10\pm0.02$ & $0.26\pm0.05$ & $0.17\pm0.02$ \\[2mm]
    $c_{L/G}$ (GeV$^2$) & $0$ & $0.20\pm0.08$ & $0.54\pm0.17$ & $0.36\pm0.08$ \\[2mm]
    $c_{q/{\rm NP}}$ (GeV$^{-2}$) & $0.87\pm0.13$ & $1.22\pm0.04$ & $0.91\pm0.05$ & $1.09\pm0.05$ \\[2mm]
    $c_\Reg$ (GeV$^{-2}$) & $6.7\pm0.8$ & --- & $7.5\pm2.0$ & $6.2\pm0.6$ \\[2mm]
    $\lambda$ & $0.23\pm0.04$ & $0.21\pm0.02$ & $0.13\pm0.01$ & $0.17\pm0.01$ \\[2mm]
    $N_Z$ & --- & $1.56$ (fixed) & --- & $1.56\pm0.06$ \\[2mm]
    $N_H$ & --- & --- & $1.26$ (fixed) & $1.26\pm0.05$ \\[2mm]\hline
    $R(6.5\,\mathrm{GeV}^2), R(90\,\mathrm{GeV}^2)$ & $0.60,0.60$ & $0.56,0.57$ & $0.54,0.55$ & $0.55,0.56$ \\ \hline
  \end{tabular}
  \caption{The values of the free parameters obtained in the fits to preliminary ZEUS \cite{ZEUSLPS,ZEUSMX} and H1 \cite{H1data} $F_2^{D(3)}$ data with a gluon distribution of the proton proportional to $\xPom^{-\lambda}$ \eqref{eq:lambda}.  The last row $R(Q^2)$, defined in \eqref{eq:glufrac}, gives the fraction of the Pomeron's (plus Reggeon's) momentum carried by gluons at $\xPom=0.003$.}
  \label{tab:dummyG}
\end{table}

\begin{figure}
  \centering
  \begin{minipage}{0.49\textwidth}
    \hspace{0.5\textwidth}(a)\\
    \includegraphics[width=\textwidth]{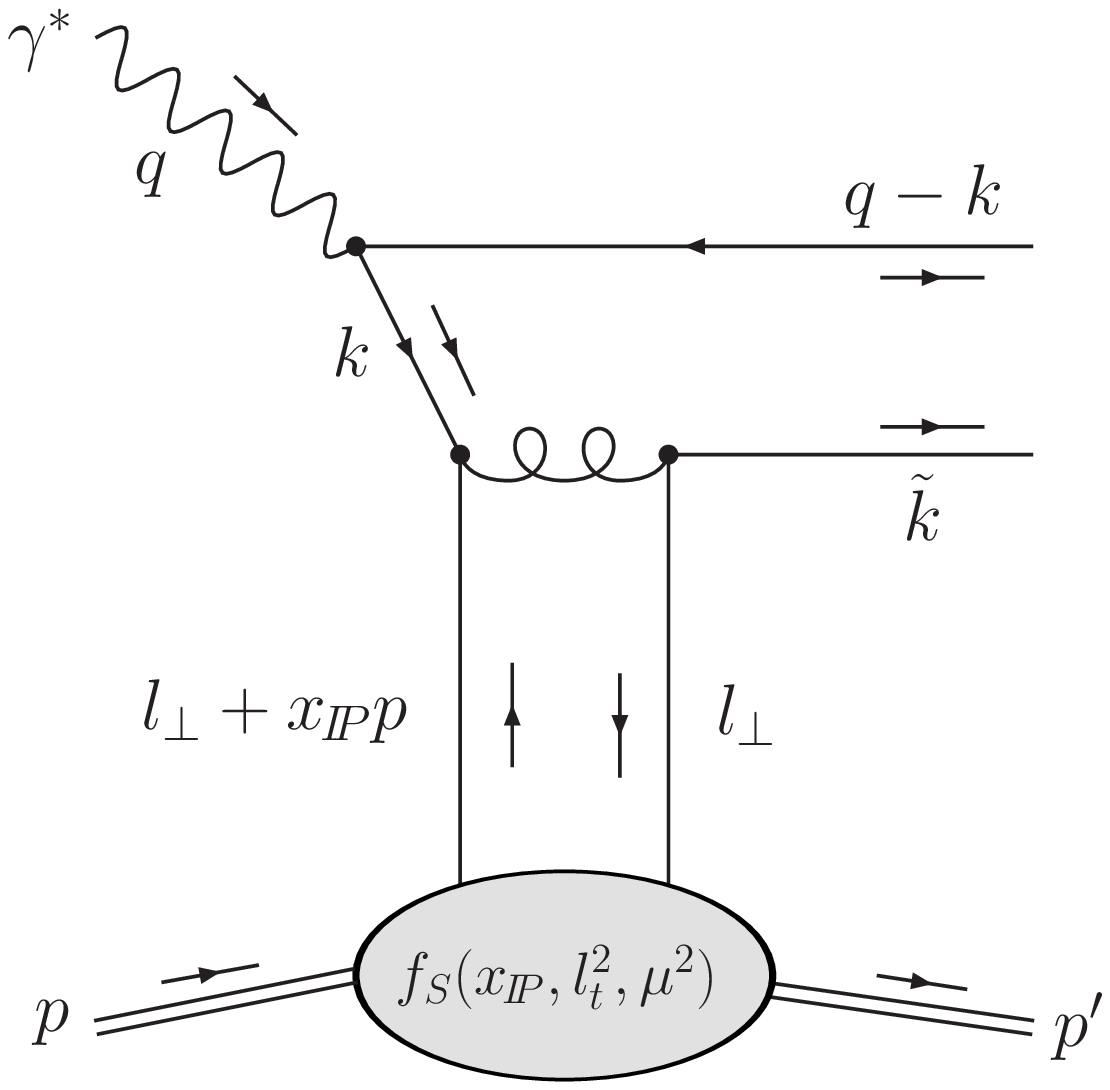}
  \end{minipage}
  \begin{minipage}{0.49\textwidth}
    \hspace{0.5\textwidth}(b)\\
    \includegraphics[width=\textwidth]{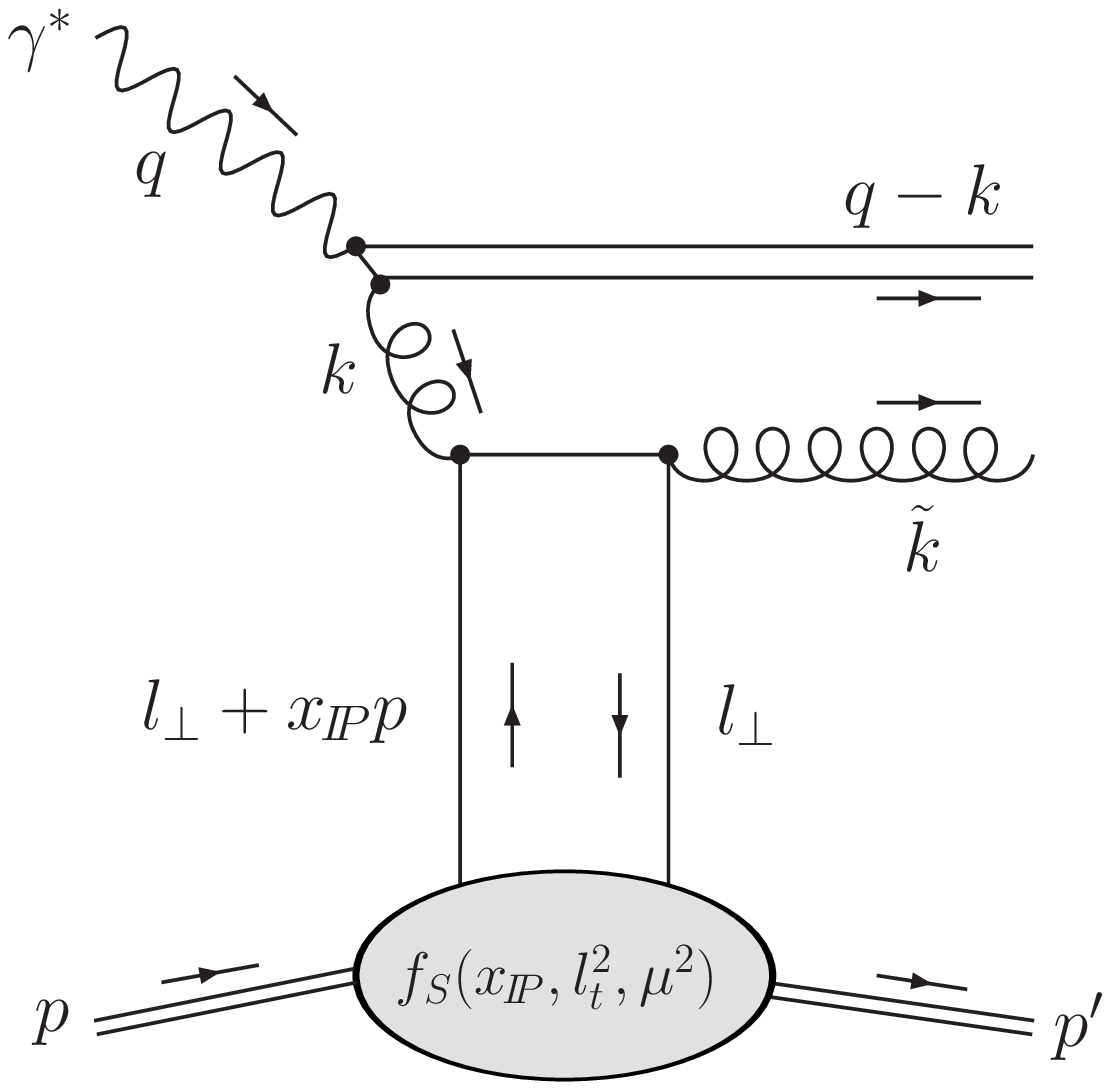}
  \end{minipage}
  \caption{(a) Quark dipole and (b) effective gluon dipole interacting with the proton via a perturbative Pomeron composed of two $t$-channel sea quarks.  Here, $f_S(\xPom,l_t^2,\mu^2)$ is the unintegrated sea quark distribution of the proton.}
  \label{fig:dipoleS}
\end{figure}

These fits to the DDIS data imply that the growth of $F_2^{D(3)}$ with decreasing $\xPom$ comes from a gluon distribution which grows as $\xPom^{-\lambda}$ with $\lambda \simeq 0.17$.   On the other hand, at low scales $\mu\sim Q_0\sim 1$ GeV, which are dominant due to the $1/\mu^4$ factor in the Pomeron flux factor \eqref{eq:PpomfluxG}, the gluon distribution of the proton obtained from global analyses of DIS and related data is valence-like, or even negative, at small $x$, while the sea quark distribution grows as a negative power of $x$ with decreasing $x$.  Therefore, in order to describe the DDIS data we are forced to introduce a Pomeron comprised of two $t$-channel sea quarks, illustrated in Fig.~\ref{fig:f2d3pom}(b).  In analogy to the flux factor \eqref{eq:PpomfluxG} for the two-gluon component, we therefore define a perturbative Pomeron flux factor for this two-quark component ($\Pom=S$) and for the interference term ($\Pom=GS$):
\begin{align}
  f_{\Pom=S}(\xPom;\mu^2) &= \frac{1}{\xPom} \left[\,\frac{\alpha_S(\mu^2)}{\mu^2}\;\xPom\,S(\xPom,\mu^2)\,\right]^2, \label{eq:PpomfluxS} \\
  f_{\Pom=GS}(\xPom;\mu^2) &= \frac{1}{\xPom} \left[\,\frac{\alpha_S(\mu^2)}{\mu^2}\,\right]^2\;2\,\xPom\,g(\xPom,\mu^2)\;\xPom\,S(\xPom,\mu^2), \label{eq:PpomfluxGS}
\end{align}
where $S(\xPom,\mu^2)\equiv 2[\bar{u}(\xPom,\mu^2)+\bar{d}(\xPom,\mu^2)+\bar{s}(\xPom,\mu^2)]$ is the integrated sea quark distribution of the proton.  Just as for the starting distributions of the two-gluon Pomeron, \eqref{eq:dpdfqpG} and \eqref{eq:dpdfgpG}, we calculate the $\beta$ dependence of the input Pomeron parton distributions for the two-quark Pomeron (and the interference contribution) from the diagrams in Fig.~\ref{fig:dipoleS} using lowest-order perturbative QCD.  Unlike for the two-gluon Pomeron, there are only two permutations of the couplings of the two $t$-channel sea quarks to the two components of the dipole.  We find\footnote{Note that $\beta \Sigma^{\Pom=S}(\beta,\mu^2;\mu^2)$ has the same $\beta$ dependence as the $F_{q\bar{q}}^T$ term in the BEKW model \cite{Bartels:1998ea}.}
\begin{align}
  \beta \Sigma^{\Pom=S}(\beta,\mu^2;\mu^2) &= c_{q/S}\,\beta\,(1-\beta), \label{eq:dpdfqpS} \\
  \beta^\prime g^{\Pom=S}(\beta^\prime,\mu^2;\mu^2) &= c_{g/S}\,(1-\beta^\prime)^2, \label{eq:dpdfgpS} \\
  \beta \Sigma^{\Pom=GS}(\beta,\mu^2;\mu^2) &= c_{q/GS}\,\beta^2\,(1-\beta), \label{eq:dpdfqpGS} \\
  \beta^\prime g^{\Pom=GS}(\beta^\prime,\mu^2;\mu^2) &= c_{g/GS}\,(1+2\beta^\prime)\,(1-\beta^\prime)^2. \label{eq:dpdfgpGS}
\end{align}
The $\beta$ dependent factors of the twist-four contribution arising from Fig.~\ref{fig:dipoleS}(a) with a longitudinally polarised photon (and the interference contribution), analogous to \eqref{eq:FLbeta}, are found to be
\begin{align}
  F_{L}^{\Pom=S}(\beta) &= c_{L/S}\;\beta^3, \\
  F_{L}^{\Pom=GS}(\beta) &= c_{L/GS}\;\beta^3\,(2\beta-1).
\end{align}
The normalisation of the interference terms between the two-gluon and the two-quark Pomerons is fixed by $c_{i/GS}=\sqrt{c_{i/G}\; c_{i/S}}$, where $i=q,g,L$; that is, the $K$-factor is fixed for the amplitude rather than for the cross section.

The results of fits with this extended model, using the MRST2001 NLO \cite{MRST2001} gluon and sea quark distributions of the proton, are shown in Table \ref{tab:MRST}.  We set $\xPom\,g(\xPom,\mu^2)=0$ if it is negative.  Again, good fits are obtained whether fitting ZEUS and H1 data separately or all together.  However, the fit with only H1 data is dramatically different from the other three fits in Table \ref{tab:MRST}, with a much larger two-gluon Pomeron contribution compared to the other three, which are dominated by the two-quark Pomeron.  This difference can be traced to the flatter $\xPom$ dependence of the H1 data compared to the ZEUS data (see Table \ref{tab:dummyG}).  Note that some parameters in Table \ref{tab:MRST} are consistent with zero, indicating some redundancy in this extended model.
\begin{table}
  \centering
  \begin{tabular}[c]{lcccc}
    \hline
    Data sets fitted & ZEUS LPS & ZEUS $M_X$ & H1 & ZEUS + H1 \\
    Number of points & 69 & 121 & 214 & 404 \\
    \hline
    $\chisq$ & $0.79$ & $0.96$ & $0.71$ & $1.14$ \\[2mm]\hline
    $c_{q/G}$ (GeV$^2$) & $0.001\pm0.053$ & $0.018\pm0.023$ & $0.36\pm0.06$ & $0.18\pm0.04$ \\[2mm]
    $c_{g/G}$ (GeV$^2$) & $0$ & $0$ & $0.37\pm0.02$ & $0$ \\[2mm]
    $c_{L/G}$ (GeV$^2$) & $0.21\pm1.48$ & $0.050\pm0.033$ & $0.14\pm0.03$ & $0.064\pm0.024$ \\[2mm]
    $c_{q/S}$ (GeV$^2$) & $0.97\pm0.40$ & $0.49\pm0.10$ & $1.06\pm0.13$ & $0.58\pm0.07$ \\[2mm]
    $c_{g/S}$ (GeV$^2$) & $1.23\pm0.21$ & $1.23\pm0.07$ & $0$ & $1.31\pm0.07$ \\[2mm]
    $c_{L/S}$ (GeV$^2$) & $0.41\pm0.28$ & $0.21\pm0.09$ & $0$ & $0.11\pm0.05$ \\[2mm]
    $c_{q/{\rm NP}}$ (GeV$^{-2}$) & $0.79\pm0.22$ & $1.16\pm0.08$ & $0.09\pm0.11$ & $0.92\pm0.07$ \\[2mm]
    $c_\Reg$ (GeV$^{-2}$) & $6.6\pm0.7$ & --- & $8.4\pm1.8$ & $6.4\pm0.5$ \\[2mm]
    $N_Z$ & --- & $1.54$ (fixed) & --- & $1.54\pm0.06$ \\[2mm]
    $N_H$ & --- & --- & $1.24$ (fixed) & $1.24\pm0.04$ \\[2mm]\hline
    $R(6.5\,\mathrm{GeV}^2), R(90\,\mathrm{GeV}^2)$ & 0.57, 0.58 & 0.57, 0.59 & 0.60, 0.66 & 0.57, 0.57 \\ \hline
  \end{tabular}
  \caption{The values of the free parameters obtained in the fits to ZEUS \cite{ZEUSLPS,ZEUSMX} and H1 \cite{H1data} $F_2^{D(3)}$ data with MRST2001 NLO \cite{MRST2001} gluon and sea quark distributions of the proton.  The last row $R(Q^2)$, defined in \eqref{eq:glufrac}, gives the fraction of the Pomeron's (plus Reggeon's) momentum carried by gluons at $\xPom=0.003$.}
  \label{tab:MRST}
\end{table}

From these fits to $F_2^{D(3)}$, we can extract diffractive parton distributions, $a^D(\xPom,\beta,Q^2) = \beta\Sigma^D(\xPom,\beta,Q^2)$ or $\beta g^D(\xPom,\beta,Q^2)$, from the three leading-twist contributions to \eqref{eq:F2D3sum}:
\begin{multline} \label{eq:dpdfs}
  a^D(\xPom,\beta,Q^2) = \sum_{\Pom=G,S,GS}\left(\int_{Q_0^2}^{Q^2}\!\dif\mu^2\;f_{\Pom}(\xPom;\mu^2)\;a^\Pom(\beta,Q^2;\mu^2)\right) \\+ f_{\Pom={\rm NP}}(\xPom)\,a^{\Pom={\rm NP}}(\beta,Q^2;Q_0^2) + c_\Reg\,f_\Reg(\xPom)\,a^\Reg(\beta,Q^2).
\end{multline}
\begin{figure}
  (a)
  \begin{center}
    \includegraphics[width=0.9\textwidth,clip]{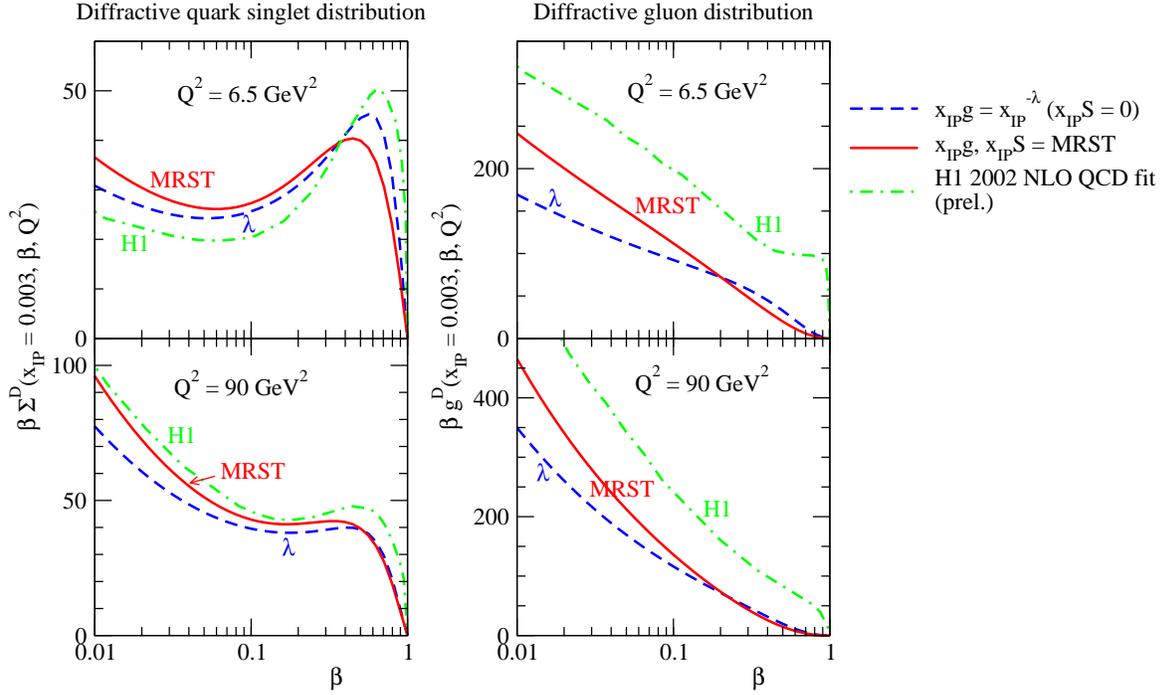}
  \end{center}
  (b)
  \begin{center}
    \includegraphics[width=0.9\textwidth,clip]{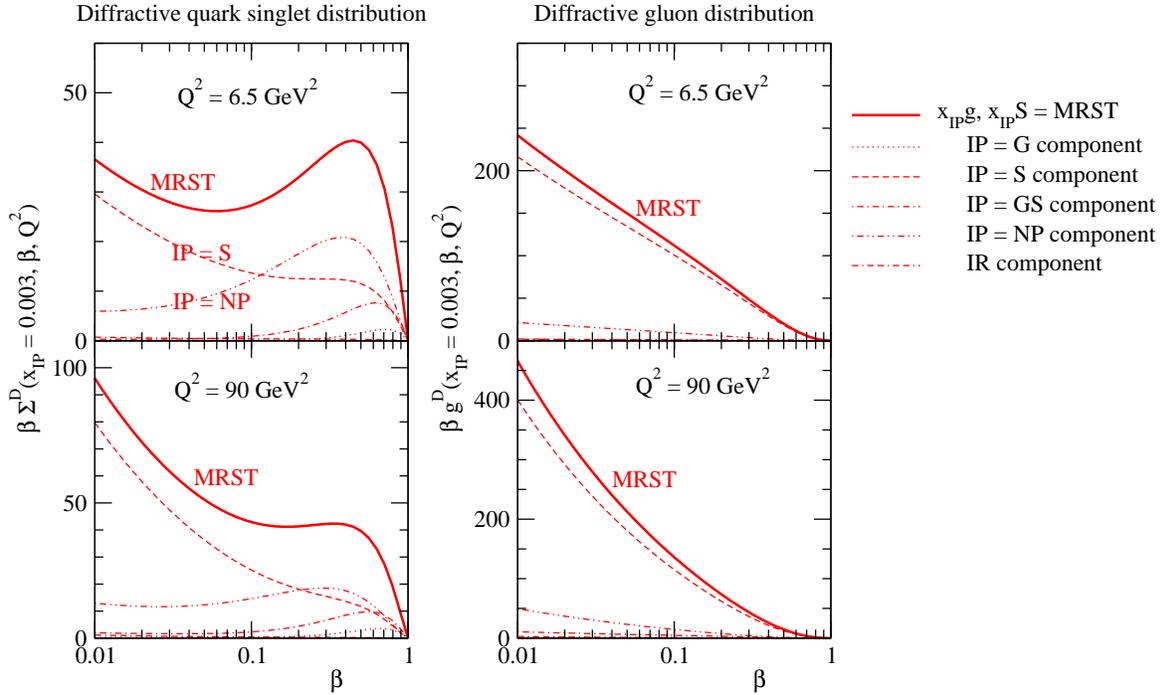}
  \end{center}
  \caption{(a) The curves labelled `$\lambda$' and `MRST' show the diffractive parton distributions extracted from the fits in Tables \ref{tab:dummyG} and \ref{tab:MRST} to the combined preliminary ZEUS \cite{ZEUSLPS,ZEUSMX} and H1 \cite{H1data} data (compared to those obtained by H1 \cite{H1data}).  (b) The breakdown of the five separate components of \eqref{eq:dpdfs} for the `MRST' fit to the combined data sets.}
  \label{fig:dpdfs}
\end{figure}
The diffractive parton distributions calculated using \eqref{eq:dpdfs} are plotted for the fits to the combined ZEUS and H1 data sets of Tables \ref{tab:dummyG} (`$\lambda$') and \ref{tab:MRST} (`MRST') in Fig.~\ref{fig:dpdfs}(a) for $\xPom=0.003$ and $Q^2=6.5,90$ GeV$^2$.  We also show the Pomeron parton distributions from the preliminary H1 analysis \cite{H1data} multiplied by $f_\Pom(\xPom)$ (given by \eqref{eq:NPregflux} with $\alpha_\Pom(0)=1.173$) and normalised to the ZEUS LPS data by dividing by a factor 1.26 (from Table \ref{tab:dummyG}).  The eight different fits of Tables \ref{tab:dummyG} and \ref{tab:MRST} are found to give similar diffractive parton distributions, especially at the higher $Q^2$ value, with the possible exception of the `MRST' fit to only H1 data.  In Fig.~\ref{fig:dpdfs}(b) we show the breakdown of the five separate components of \eqref{eq:dpdfs} for the `MRST' fit to the combined data sets.  Note the large contribution from the two-quark component of the Pomeron.

From Fig.~\ref{fig:dpdfs}, the diffractive quark singlet distribution obtained by H1 \cite{H1data} has a slightly steeper $Q^2$ dependence than the fits presented here, and hence H1 obtain a larger diffractive gluon distribution.  In addition, the smaller value of $\alpha_S(M_Z^2)$ used by H1 also enlarges their gluon density.\footnote{In the preliminary H1 analysis \cite{H1data}, $\Lambda_{\rm QCD}=0.2$ GeV for 4 flavours, corresponding to $\alpha_S(M_Z^2)=0.1085$, whereas we take $\alpha_S(M_Z^2)=0.1190$ from the MRST2001 NLO parton set \cite{MRST2001}; cf.~the world average, $\alpha_S(M_Z^2)=0.1187(20)$, from the PDG \cite{PDG2004}.}  In our analysis, all the input Pomeron parton distributions vanish as either $(1-\beta)$ or $(1-\beta)^2$ as $\beta\to1$.  As $\beta\to1$, the only non-zero contribution to $F_2^{D(3)}$ comes from a twist-four component arising from longitudinally polarised photons.  This contribution was not included in the H1 analysis \cite{H1data}, and hence rather large diffractive parton distributions were obtained by H1 for $\beta$ close to 1, with an unphysical `bump' in the diffractive gluon distribution (see Fig.~\ref{fig:dpdfs}).

We define the fraction of the Pomeron's (plus Reggeon's) momentum carried by gluons at $\xPom = 0.003$ as
\begin{equation} \label{eq:glufrac}
  R(Q^2) \equiv \frac{\int_{0.01}^1\!\dif\beta\;\beta g^D(\xPom=0.003,\beta,Q^2)}{\int_{0.01}^1\!\dif\beta\;\left[\beta\Sigma^D(\xPom=0.003,\beta,Q^2)+\beta g^D(\xPom=0.003,\beta,Q^2)\right]},
\end{equation}
which is given for $Q^2$ values of $6.5$ and $90$ GeV$^2$ in the last rows of Tables \ref{tab:dummyG} and \ref{tab:MRST}.  The gluon momentum fraction, $R(Q^2)$, is consistently 55--60\% and is almost independent of $Q^2$.  Taking the same $\alpha_S(M_Z^2)$ as in the preliminary H1 analysis would increase this value to $\approx65\%$, compared to the value found by H1 of $75\pm15$\% \cite{H1data}.

Note, from Fig.~\ref{fig:dummyG}, that the perturbative Pomeron contribution to $F_2^{D(3)}$ (from scales $\mu>Q_0=1$ GeV) is not small; as a rule it is more than half the total contribution.  The perturbative contribution is even stronger for the `MRST' fits presented in Table \ref{tab:MRST}.  The comparison of the separate fits to the ZEUS and H1 data presented in Table \ref{tab:MRST} demonstrates that there is a strong correlation between the pairs of parameters $c_{i/G}$ and $c_{i/S}$, where $i=q,g,L$.  That is, with the present accuracy of the data, it is hard to distinguish between partons which originate from the two-gluon and two-quark components of the Pomeron.\footnote{The combined analysis of DDIS data with a more exclusive diffractive process, such as diffractive $J/\psi$ production at HERA, which is sensitive to the two-gluon component of the Pomeron, may help to resolve this problem.}  Nevertheless, the final diffractive parton distributions are similar for the different fits.  This stability increases confidence in these distributions, so that they can be used in the description of other diffractive processes at HERA and the Tevatron.  Of course, we must include the probability that the rapidity gap survives the soft rescattering of the colliding hadrons or `hadron-like' states \cite{Khoze:2000wk}.

An advantage of describing the diffractive DIS data using an approach with an explicit integral over the Pomeron scale $\mu$ is the possibility to use the results to evaluate the absorptive corrections, $\Delta F_2^\mathrm{abs}(\xB,Q^2;\mu^2)$, to the inclusive structure function $F_2(\xB,Q^2)$.  These corrections arise during the DGLAP $\ln q^2$ evolution at each point $q^2=\mu^2$.  With this approach the AGK cutting rules \cite{AGK} give
\begin{equation}
  \label{eq:agk}
  \Delta F_2^\mathrm{abs}(\xB,Q^2;\mu^2) \simeq - F_2^D(\xB,Q^2;\mu^2),
\end{equation}
where $F_2^D(\xB,Q^2;\mu^2)$ is the contribution to the diffractive structure function $F_2^{D(3)}$ (integrated over $\xPom$) which originates from a perturbative component of the Pomeron of size $1/\mu$.  In a separate paper \cite{MRW2}, we use this property to incorporate the absorptive corrections in a parton analysis of inclusive DIS data.

\section*{Acknowledgements}

We thank Michele Arneodo and G\"unter Wolf for useful discussions.  ADM thanks the Leverhulme Trust for an Emeritus Fellowship.  This work was supported by the UK Particle Physics and Astronomy Research Council, by the Federal Program of the Russian Ministry of Industry, Science and Technology (grant SS-1124.2003.2), by the Russian Fund for Fundamental Research (grant 04-02-16073), and by a Royal Society Joint Project Grant with the former Soviet Union.

\end{document}